

\magnification = 1200
\tolerance 1000
\pretolerance 1000
\baselineskip 20pt
\overfullrule = 0pt
\hsize=16.3truecm

\vskip 0.5truecm

\centerline {\bf  P.M.A. Cook, V.C. Hui$\dagger$ and C.J. Lambert, }

\centerline {\bf School of Physics and Materials}

\centerline {\bf Lancaster University}

\centerline {\bf Lancaster   LA1 4YB}

\centerline {\bf U.K.}

\vskip 0.5truecm

\centerline {\bf$\dagger$ Centre for Theoretical Studies }
\centerline {\bf D.R.A Malvern  WR1 3PS   UK }

\vskip 0.5truecm

\noindent
ABSTRACT.
A new class of nano-structure devices is suggested, based on interference
from the order parameter phase gradient of a single superconductor (S)
in contact with a single normal  metallic lead (N).
By solving the Bogoliubov -  de Gennes equation in two dimensions, it is
demonstrated that the electrical conductance of a normal conductor of
width $M$ in contact with a superconductor will oscillate as the phase
gradient $v$ at 90$^0$ to the interface is increased. This effect is
enhanced by the presence of a Schottky barrier at the interface and
is also present in $N-S-N$ structures.


\vfil\eject
\noindent

Following the pioneering work of Spivak and Khmel'nitskii[1],
it has long been recognised that the conductance of a phase coherent normal
structure with two superconducting inclusions should oscillate
 with the phase difference $\phi$ between the
 order parameters of the inclusions[2-4]. While the details of this effect
is still a subject of discussion[5-9], the underlying principle
 behind such Andreev interferometers
 is clear; the wavefunction of a quasi-particle which
  Andreev reflects at a normal-superconducting (N-S)
 interface aquires the local phase of the order parameter. Therefore
 if two such interfaces are present, interference between partial
 waves reflecting from the separate interfaces
 yields an Andreev reflection coefficient $R_a$ of the form
 $R_a=A+B\cos\phi + \dots$, where the dots indicate the presence of
  higher harmonics, details of which depend on the geometry and underlying
  disorder of the sample. If $\mu$ is the common condensate
chemical potential of the inclusions and $\mu_1=\mu + eV$ the
chemical  potential of an
external normal reservoir of electrons, the current $I$ flowing from the
reservoir
into the superconductor is $I=(2e^2/h)G_{\rm NS}V$, where
$G_{\rm NS}=2R_a$ is the BTK boundary conductance[10].
Consequently through  measurements of the electrical conductance,
it has been possible to confirm
the existence of Andreev interferometers in nano-structure devices
formed from tunnel junctions[11]
and metallic interfaces[12,13].

The aim of this Letter is to suggest a new class of devices based on
interference from the order parameter phase gradient of a single
superconductor. Such Andreev phase gradiometers should be easier to
construct than a structure containing two interfaces and may open up the
possibility of measuring unambiguous Andreev interference effects in phase
coherent normal-semiconductor structures. Four examples of phase gradiometers
are shown in figure 1. In examples A,B,C, which are studied in detail below,
 the measured current $I$
flows vertically from
a normal, crystalline
external lead at the bottom of the figure to another at the top.
The leads are connected to normal reservoirs at potentials $V_1$ and $V_2$
respectively and we shall compute the total electrical conductance
$G=(h/2e^2)I/(V_1-V_2)$ in the presence of a superconducting order
parameter of the form $\Delta(\underline r)=\Delta_0{\rm exp}(ivx)$,
with the {\t x}-axis chosen to be horizontal. In practice one can envisage
producing a phase gradient $v$ of this kind, at 90$^o$ to the measured
current $I$ by applying a control current from left to right.
Example A of figure 1 shows a clean superconductor in contact with metallic
leads. Example B shows a (N-I-S-N)
structure with a single insulating (I) barrier at one interface. Example
C shows a N-I-N-S-N structure, with a metallic region between the insulating
barrier and the superconductor.

The key principle we wish to establish is that the conductance of these
devices is an oscillatory function of the phase gradient $v$.
One expects this behaviour, because the order parameter acts as a
complex, off-diagonal scattering potential and therefore for a two-dimensional
N-S interface  of finite width $M$, scattering matrix elements will be
sensitive to the total phase change $Mv$. Consequently
transport coefficients should be oscillatory functions of $v$, with
period $\overline v=2\pi/M$.
A further aim of this Letter is to determine which if any of the examples
shown in figure 1 yield an
oscillation which is a finite fraction of the overall conductance.

The central quantity needed to compute transport properties
of a phase coherent sample possessing a Hamiltonian $H$ and
 connected to external current carrying leads,
is the quantum mechanical scattering matrix $s(E,H)$, with sub-matrices
$s^{\alpha,\beta}_{L,L'}(E,H)$, which describe the scattering of  excitations
of energy $E$
 from all incoming $\beta$ channels
of  lead
 $L'$ to all outgoing $\alpha$ channels of lead $L$ (where
 $\alpha , \beta $ = +1 for particles and -1 for holes).
 From a knowledge of $s(E,H)$, a matrix of reflection
and transmission coefficients can be constructed
$ P^{\alpha \beta}_{L,L'}(E) =
{\rm Trace}\{s^{\alpha \beta}_{L,L'}(s^{\alpha \beta}_{L,L'})
^\dagger \} $, in terms of which  the zero temperature,
two probe electrical conductance, in units of $2e^2/h$,
can be written[14,15],
$$G=T_0+T_a + {{2(R_a R'_a -T_a T'_a)}\over {R_a+R'_a+T_a+T'_a}}
\eqno{(1)}.$$
The coefficients $R_0= P^{++}_{L,L}(0)$, $T_0= P^{++}_{L',L}(0)$
($R_a= P^{-+}_{L,L}(0)$, $T_a= P^{-+}_{L',L}(0)$) are
probabilities for normal (Andreev)
reflection and transmission of quasi-particles from the lower
reservoir $L$, while
$R'_0, T'_0$ ($R'_a, T'_a$) are corresponding probabilities for
quasi-particles from the upper reservoir $L'$. In the presence of $N$ open
channels
per lead, these satisfy
$R_0+T_0+R_a+T_a=R'_0+T'_0+R'_a+T'_a=N$
and
$T_0+T_a=T'_0+T'_a .$
In the limit of negligible quasi-particle transmission, the resistance
 reduces to a simple sum  of
two BTK boundary resistances,
$$G^{-1}=2/R_a + 2/R'_a\eqno{(2)}$$
associated with Andreev reflection into the separate
reservoirs.

Given the spatial form of the superconducting order parameter
$\Delta(\underline r)$ the scattering
matrix can be computed by solving the Bogoliubov - de Gennes equation,
as outlined in [15].  In what follows, we present the results of
detailed numerical simulations of a two dimensional tight binding system,
described by a
Bogoliubov - de Gennes operator of the form
$$H =
\left(\matrix{H_0
& \Delta \cr
\Delta^*
& -H_0^* }\right) \sp .$$
In this equation $H_0$  is a nearest neighbour tight binding model
on a square lattice, with off-diagonal hopping elements of value $-\gamma$
and $\Delta$ a diagonal order parameter matrix.
 The scattering region
  is chosen to be  $M$ sites wide and is connected to
external leads of width $M$, as shown in figures 1A to 1C.
For convenience we make the choice $\gamma=1$ and
throughout the whole structure, except in a barrier region,
the diagonal elements of $H_0$ are set to $10^{-3}$.
 By choosing a value which is close to, but
not identically
zero one obtains a normal
host material close to half-filling, while avoiding a discontinuity in the
number
of open channels at $E=0$. For structures B and C, within the region
occupied by the barrier,
diagonal elements of $H_0$
are set to $3\gamma$.
Finally within the superconductor, the magnitude $\Delta_0$ of the order
parameter is chosen to be $\Delta_0=0.5\gamma$ and since $\mu\approx 4\gamma$,
$\Delta_0/\mu\approx10^{-1}$, which is typical of a cuprate superconductor.
In what follows, for each structure and a given choice of $v$,
 the  scattering matrix at $E=0$ is obtained
numerically, using a transfer matrix
 technique outlined in appendix 2 of reference[15]. For a lattice constant
 $a$, the Landau critical velocity of a such a homogeneous superconductor
 is $v^*=\Delta_0/(a\gamma)$. In what follows we choose $a=1$ and therefore
 $v^*=\Delta_0=0.5$.

For $M=30$ and a superconductor of length $M'=20$, figure 2 shows the variation
of the various scattering coefficients, along with the electrical conductance
$G$. Since this structure is symmetric about a horizontal line passing through
the centre of the superconductor, all scattering coefficients from lead
1 are identical to those from lead 2
and the conductance formula (1)
reduces to $G = T_{o} + R_{a} $. Although careful inspection of
the curves of figure 2 reveal a periodic modulation with $v$, the effect is
clearly negligible. We have carried out simulations for a range of $M$ and
$M'$ and have found only a negligible effect for all structures of type A.

Figure 3 shows results for three systems of type B, with widths $M=15$, $30$
and $45$, a potential barrier of length $5$ and a long superconductor of
length $M'=150$. The latter is chosen to yield negligible transmission
through the device, so that equation (2) provides a good approximation to
$G$. Furthermore the device is now asymmetric and the overall resistance
is dominated by the boundary conductance $2R_a$ of the lower interface.
It is clear from this figure that the period of oscillation
is inversely proportional to the width of the device and that
by introducing a barrier,
the relative size of the effect is increased. This enhancement is reminiscent
of the increase of zero bias anomalies through
 the presence of a Schottky barrier[16-21].

We have performed numerical simulations of a variety of structures and in
all cases find that the presence of a barrier at the interface
enhances this effect. For a simple structure such as that modelled in figure 3,
a familiar Fraunhoffer diffraction pattern is obtained,
which can be understood by
examining the overlap between an outgoing plane wave with a transverse
wavevector
shifted by the continuous variable $v$ and the discrete number of
allowed wavevectors defining open channels in the external leads.
As an example of a more complex system, figure 4 shows results for structure C
of figure 1. In this case, the system is of width $M=30$, the superconductor
of length $M'=9$, the insulating
barrier of length $5$ and the normal, crystalline,
metallic region separating the barrier from the superconductor is of
length $6$. For this structure resonances associated with multiple scattering
within
the normal region yield a more complex interference pattern, which is reflected
in the non-trivial variation of $G$ with $v$.

The aim of these simulations has been to demonstrate that an oscillatory
dependence of $G$ on the phase gradient $v$ is a generic feature of hybrid
superconducting structures and to
identify systems for which the effect is non-negligible.
To avoid relying on approximate analytical solutions, we have chosen
to produce results
based on exact solutions of the Bogoliubov - de Gennes equation in
two dimensions.
The simulation leading to figure 3 reveals that this effect is enhanced by
the presence of a Schottky barrier at the interface and therefore it should
be possible to observe these oscillations
 in semiconductor-superconductor structures.
Figure 3 also shows that the effect is present in
the BTK boundary conductance and therefore the superconductor itself
can be used as one of the external reservoirs. From the point
of view of constructing the simplest possible
experiment, this leads us to suggest the structure sketched in example
D of figure 1 as a possible candidate.
In this example a superconducting loop is connected via a Schottky
barrier to a normal lead and the boundary conductance $2R_a =
(h/2e)I/(\mu_1-\mu)$ is measured as a function of the magnetic field
through the loop. Since the latter produces a surface screening current
and therefore a phase gradient on the N-S interface, the boundary conductance
will oscillate on a  field scale which depends on the width $M$ of the normal
lead and can therefore be chosen to be distinct from that
that of a flux quantum through the loop. Finally to observe this effect,
one notes that
the period $\overline v=2\pi/M$ must be smaller than the critical velocity
$v^*$, which implies that the width $M$ must be greater than the
zero temperature coherence length of the superconductor.

\vfil\eject

\vskip 0.5truecm
\noindent
{\bf Acknowledgements.}
This work is supported by the SERC, the EC, the MOD and NATO.
\vskip 2.0truecm
\noindent
$\underline {\hbox {\bf Figure Captions.}}$
\vskip 0.5truecm
\noindent
{\bf Figure 1.}
Sketches A,B and C show the three distinct  structures used in the simulations.
Sketch D shows the simplest possible candidate for an experimentally realisable
Andreev phase gradiometer.

\noindent
{\bf Figure 2.}
This shows the variation with $v$
of G , $R_{o}$ , $R_{a}$ , $T_{o}$ and $T_{a}$ with V , for a
system of type A. For convenience all quantities are divided by the
number $N$ of open channels.

\noindent
{\bf Figure 3.}
This shows the variation with $v$ of G and $R_{a}$ for a system of type B
with 3 different widths, $R_{a}$ is the Andreev
reflection co-efficient for the superconducting interface containing the
barrier. All quantities are divided by the number of open channels.

\noindent
{\bf Figure 4.}
This shows variation with $v$ of $G$ and $R_{a}$  for a system of
type C . Both quantities are divided by the number of open channels.

\vfill\eject
\noindent
$\underline {\hbox {\bf References.}}$

\item{1.} B.Z. Spivak and D.E. Khmel'nitskii, JETP Lett, {\bf35} 413 (1982).

\item{2.} B.L. Al'tshuler and B.Z. Spivak, Sov. Phys. JETP {\bf 65}, 343
(1987).

\item{3.} H. Nakano and H. Takayanagi, Sol. St. Comm. {\bf 80} 997 (1991)

\item {4.} S. Takagi, Sol. St. Comm. {\bf 81} 579 (1992)

\item{5.}  C.J. Lambert  J.Phys. Condensed Matt. {\bf 5} 707 (1993)

\item {6.} V.C. Hui and C.J. Lambert, Europhys. Lett. {\bf 23} 203 (1993)

\item{7.} C.J. Lambert and S.J. Robinson, Physica {\bf B194-196} 1641 (1994)

\item{8.} F.W. Hekking and Yu. Nazarov, Phys. Rev. {\bf B49} 6847 (1994).

\item{9.} V.C. Hui and C.J. Lambert, Physica {\bf B194-196} 1673 (1994)

\item{10.} G.E. Blonder, M. Tinkham and T.M. Klapwijk, Phys. Rev. B. {\bf 25}
4515 (1982).

\item{11.} H. Pothier, D. Esteve and M.H. Devoret, preprint

\item{12.} J. Price, private communication and proceedings of
the A.P.S. March Meeting, Pittsburgh (1994).

\item{13.} V.T. Petrashov, Proceedings of the ICTP Workshop on Quantum
Dynamics of Sub-micron Strutures, Trieste (1994), to be published
Physica {\bf B} (1994)

\item{14.} C.J. Lambert, J. Phys.: Condensed Matter, {\bf 3} 6579 (1991)

\item{15.} C.J. Lambert, V.C. Hui and S.J. Robinson, J.Phys.: Condens. Matter,
{\bf 5} 4187 (1993)

\item{16.} A.V. Zaitsev, JETP Lett. {\bf 51} 41 (1990)

\item{17.} A. Kastalasky, A.W. Kleinsasser, L.H. Greene, R. Bhat and
J.P. Harbison, Phys. Rev. Lett. {\bf 67} 3026 (1991)

\item{18.} B.J. van Wees, P. de Vries, P. Magnee and T.M. Klapwijk,
Phys. Rev. Lett. {\bf 69} 510 (1992)

\item{19.} A.F. Volkov and T.M. Klapwijk, Phys. Lett. {\bf A168} 217 (1992)

\item{20.} A.F. Volkov, A.V. Zaitsev and T.M. Klapwijk, Physica  {\bf C210}
217 (1993)

\item{21.} I.K. Marmakos, C.W.J. Beenakker and R.A. Jalabert,
 Phys. Rev.{\bf B48} 2811 (1993)

\end